# An inverse Faraday effect through linear polarized light

Xingyu Yang, Ye Mou, Homero Zapata, Benoît Reynier, Bruno Gallas, and Mathieu Mivelle*

Sorbonne Université, CNRS, Institut des NanoSciences de Paris, INSP, F-75005 Paris, France
*Corresponding author: mathieu.mivelle@sorbonne-universite.fr

**ABSTRACT**

The inverse Faraday effect (IFE) allows the generation of magnetic fields by optical excitation only. Since its discovery in the 60s, it was believed that only an elliptical or circular polarization could magnetize matter by this magneto-optical phenomenon. Here, we demonstrate the generation of an IFE via a linear polarization of light. This new physical concept results from the local manipulation of light by a plasmonic nano-antenna. We demonstrate that a gold nanorod excited by a linear polarization generates a non-zero magnetic field by IFE when the incident polarization of the light is not parallel to the long axis of the rod. We show that this dissymmetry generates hot spots of local non-vanishing spin densities (local elliptical polarization state), introducing the concept of super circular light, allowing this magnetization. Moreover, by varying the angle of the incident linear polarization with respect to the nano-antenna, we demonstrate the on-demand flipping of the magnetic field orientation. Finally, this linear IFE generates a stationary magnetic field 25 times stronger than what a gold nanoparticle produces when excited by a circular polarization and via a classical IFE. The creation of stationary magnetic fields by IFE in a plasmonic nanostructure is nowadays the only technique allowing the creation of ultra-short, intense magnetic field pulses at the nanoscale. Thus, it finds applications in the ultrafast control of magnetic domains with applications not only in data storage technologies but also in research fields such as magnetic trapping, magnetic skyrmion, magnetic circular dichroism, to spin control, spin precession, spin currents, and spin waves, among others. Therefore, through linear IFE, the ultrafast manipulation of magnetic processes for the same excitation polarization, the same energy density, and the same heat generation, simply by changing the angle of the incident polarization on the plasmonic nano-antenna is a game-changer in the physics of ultrafast opto-magnetism.

## I. INTRODUCTION

In physics, the Faraday effect is a magneto-optical phenomenon that describes the interaction between light and a magnetic field in a material [1-3]. In particular, it describes the rotation of the light polarization in the material as a function of the amplitude of an external magnetic field oriented in the direction of light propagation. Conversely, the inverse Faraday effect (IFE) refers to how light can magnetize matter. This phenomenon, discovered in the 60s [1-3] and described in different materials, dielectric or metallic, is the result of non-linear forces that light applies to the material's electrons [4-7]. Fig. 1 shows the principle of the IFE. A circularly polarized electromagnetic wave is launched on a material generating, in turn, a magnetization **M** of the latter. This magnetization is proportional to the vector product **E**($\omega$)x**E**\*($\omega$), with **E** the optical electric field oscillating at the angular frequency $\omega$, the star * represents the complex conjugate and bold characters are vectors. This expression shows that only a circularly or elliptically polarized light will generate this magnetization. In particular, a right-circular polarization will create a magnetization in the direction of light propagation. In contrast, a left-circular polarization will do so in the direction opposite to the propagation [Fig. 1].

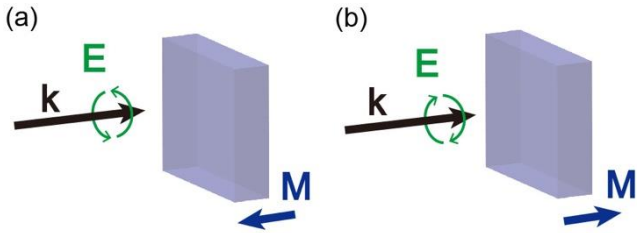

FIG. 1. Illustration of the inverse Faraday effect. a) a left circular polarization incident on a material induces a magnetization opposite to the direction of light propagation, while b) a right circular polarization creates this magnetization in the direction of light propagation.

In a metal, the IFE has been described by R. Hertel as the resultant of circular drift currents **J**$_d$ of the metal's free electrons set in motion by the non-linear forces mentioned above. These currents, which can be microscopic [7] or macroscopic [8], depend on the optical electric field and its divergence according to equation [9]:

$$\mathbf{J_d} = \frac{\sigma_\omega^2}{2en} \, \mathrm{Re}\left(\left(-\frac{\nabla.\mathbf{E}}{i\omega}\right) \cdot \mathbf{E}^*\right) \quad (1)$$

With e the charge of the electron (e < 0), n the charge density at rest and $\sigma_\omega$ the dynamic conductivity of the metal.

Recently, several theoretical and experimental papers have demonstrated that the enhancement of electric fields and field gradients by resonant plasmonic nanostructures allowed to generate, through IFE, strong drift currents and, therefore, strong stationary magnetic fields **B** (as opposed to the optical magnetic field **H** in this work) [10-14]. In particular, it was demonstrated that a plasmonic nano-antenna, devised by an inverse-design algorithm, generates a magnetic pulse of the order of the tesla at the nanoscale and for only a few femtoseconds, something that no other technique allows nowadays [9]. This world first opens the way to a whole new range of technical and scientific possibilities. For example, since the pioneering work of Beaurepaire et al. [15], researchers have intensively searched for ways to manipulate and study magnetization at the femtosecond timescale using femtosecond lasers [16] with the aim of developing high data storage and data processing technologies. However, manipulations of magnetic materials by ultrafast light pulses are still poorly understood [16-22]. The ability to probe and address magnetic processes and their transient mechanisms using ultrashort magnetic-field pulses will therefore benefit countless research activities: from Zeeman splitting [23], magnetic trapping [24], magnetic skyrmion [25], magneto-plasmonics [26], ultrafast magnetic modulation [27], and magnetic circular dichroism [28] to spin control [29], spin precession [30], spin currents [31], and spin waves [32]. Something that only the IFE through plasmonic structures will allow.

However, since the discovery of IFE and until today, it has been believed that only circular polarization could generate a stationary magnetic field by this physical phenomenon [2,3]. Here, we demonstrate that by means of a plasmonic nanostructure, we can manipulate the local polarization state of light and thus generate an IFE through excitation by a linear polarization, which is a whole new paradigm in physics and magneto-optical interactions. Moreover, the increase of the optical fields and field gradients by the resonant plasmonic nanostructure enables us to generate a **B**-field by this linear IFE much stronger than the one created by a classical IFE. In detail, we show that a gold nanorod illuminated by a linear polarization along its long axis allows the creation of local magnetic **B**-fields, which are, however, fully balanced over the whole photonic antenna. In contrast, we demonstrate that by breaking the symmetry between the incident linear polarization and the long axis of the nanorod, we can generate a uniform and intense **B**-field at the center of the nano-antenna, with a maximum 25 times higher than that created by a gold nanoparticle at the same wavelength and excited by a circular polarization. Furthermore, we show that this **B**-field is reversible on demand simply by changing the angle θ of the incident polarization on the nanorod, in particular by changing it from plus to minus θ with respect to the long axis of the plasmonic antenna. These novel and unique physical properties result from manipulating the local spin density (polarization state) to create local circular polarization around the antenna, generating, in turn, IFE and drift currents at the origin of the magnetization. The generation of intense, reversible on-demand magnetic fields by linearly polarized optical excitation at the nanoscale is an entirely new concept. It opens up exciting perspectives for manipulating magnetic materials such as magnetic domains, particularly at ultrafast time scales. Indeed, nowadays, the ultrafast manipulation of magnetic domain orientation is performed optically at micrometer scales using ultrafast pulses of right or left circularly polarized light. Using a plasmonic nano-antenna, allowing for the same optical power and the same thermal effect to generate and control an upward or downward oriented nanoscale **B**-field only by manipulating the angle of the linear polarization incident on the latter, is a game-changer in ultrafast magnetic domain manipulation.

## II. CLASSICAL INVERSE FARADAY EFFECT

A finite difference time domain algorithm performs all the simulations presented in the work. To illustrate the dependence of the IFE on circular polarization, Fig. 2 represents the magnetic field created by this magneto-optical interaction in a gold nanoparticle for a circular [Fig. 2(a)-(d)] and linear [Fig. 2(e)-(h)] polarization of excitation. The gold nanoparticle of diameter 210 nm, schematically represented in Figs. 2(a) and (e), was chosen to generate the strongest stationary magnetic field for excitation at a wavelength λ of 800 nm, in a medium of index n=1 and placed on a glass substrate. Although the nanosphere was not in resonance at 800 nm, the magnetic field generated is maximal at this wavelength due to the intrinsic properties of gold in terms of electronic conduction [See Fig. (S1) in Supplemental Material]. Figs. 2(b) and (f) show the normalized electric field distribution, calculated by finite difference time domain simulations, in the center of the particle and an xy transverse plane for excitation by a plane wave of intensity $10^{12}$ W/cm$^2$ propagating along the positive z-axis and for a polarization right-circular [Fig. 2(b)] or x-linear [Fig. 2(f)]. These distributions of electric fields and field gradients generated in the near-field of the nanoparticle create, via IFE and Eq. (1), in-plane drift currents azimuthally polarized inside the metal [Fig. 2(c) and (g) and symbolized by the white arrows in Figs. 2(b) and (f)]. In turn, through the Biot-Savart law, these currents $\mathbf{J}_d$ generate the magnetic field distributions along z described in Figs. 2(d) and (h). As previously reported [11,13], under circularly polarized excitation, a gold nanoparticle generates a non-zero stationary magnetic field of a few mT for this range of excitation intensity [Fig. 2(d)]. On the other hand, when excited by a linear polarization, the symmetry of the drift currents [Fig. 2(g)] does not allow the creation of a global non-zero magnetic field, as they cancel each other out. Nevertheless, and surprisingly, local **B**-fields can still exist around the nanoparticle due to these currents [Fig. 2(h)], something that has never been described before. However, the IFE requires a circular polarization to generate a non-zero global magnetic field for this geometry.

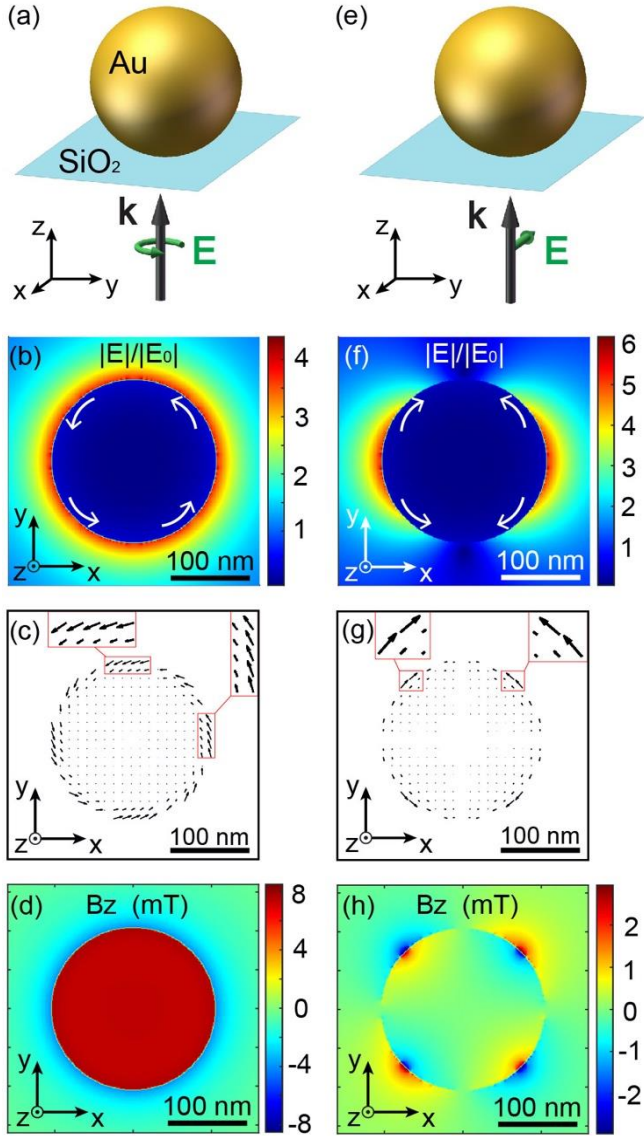

FIG. 2. The IFE in a gold nanoparticle. Generation of drift currents and stationary magnetic fields in a gold nanoparticle excited by (a-d) a circular polarization and (e-h) a linear polarization. (a, e) Schematic representation of the gold nanoparticle excited by a right-circular or x-linear polarization, respectively, at the wavelength of 800 nm and for an intensity of $10^{12}$ W/cm$^2$. For the two excitation conditions shown in (a, e), the normalized electric field distributions (b, f), the associated drift currents (c, g), and the generated stationary magnetic fields (d, h) are shown, respectively.

## III. LINEAR INVERSE FARADAY EFFECT

Gold nanospheres are one of the most popular plasmonic nanostructures for their ease of production [33] and their exceptional properties in terms of optical confinement [34,35], heat generation [36], and chemical catalysis [37], among others. Other very simple plasmonic systems that have been extensively studied are gold nanorods [38]. Because of their elongated geometry, these nano-antennas allow for manipulating the light-matter interaction in a finer way, mainly due to their polarization sensitivity to the optical electric field [39]. We use this property here to generate an IFE via the linear polarization of the light. Fig. 3 summarizes this unexpected physical phenomenon. A gold nanorod of length L=120 nm, thickness T=40 nm, width W=20nm, and placed on a glass substrate is excited by a linearly polarized plane wave of intensity $10^{12}$ W/cm$^2$, wavelength 800 nm, and propagating along the positive z-axis [Fig. 3(a) and (e)]. With these dimensions, the plasmonic nanostructure is resonant at the excitation wavelength [see Fig. (S1) in Supplemental Material]. Two linear polarizations are studied in Fig. 3. The first one [Fig. 3(a)-(d)] is oriented along y, corresponding to the long axis of the antenna [Fig. 3(a)]. The second [Fig. 3(e)-(h)] is oriented at 45° from the xy axes of the nanorod [Fig. 3(e)]. Figs. 3(b) and (f) represent the normalized electric field distributions at the center of the antenna in a transverse xy plane for these two polarizations. As shown, these distributions are extremely similar. Therefore, it is difficult to identify which pattern corresponds to which linear polarization angle. The only noticeable difference is a higher electric field enhancement for the y-polarization. On the other hand, the drift currents generated by these field distributions are very different. Indeed, while for a y-polarization, they are symmetrically distributed at the four corners of the antenna [Fig. 3(c)], for a 45° polarization, they are mainly distributed on each side of the nanorod long axis and in opposite directions [Fig. 3(g)]. Consequently, the magnetic fields generated by these currents are distributed in a very different way. In the case of polarization along the main axis of the antenna, we find a four lobes distribution around the nanorod but with opposite directions creating an overall zero magnetic field by the structure [Fig. 3(d)], similar to the case of the linearly excited gold sphere [Fig. 2(h)]. Although, and interestingly, these local **B**-fields have an amplitude 40 times higher than the case of the gold sphere excited in the same conditions. However, and remarkably, the current distribution for a 45° polarization generates a uniform magnetic field at the center of the nanorod [Fig. 3(h)], demonstrating for the first time that an IFE is possible using a linear polarization of the light, which was previously thought to be impossible. Moreover, the **B**-field has an amplitude 25 times higher than the field created by the gold sphere for the same excitation intensity and circular polarization [Fig. 2(d)].

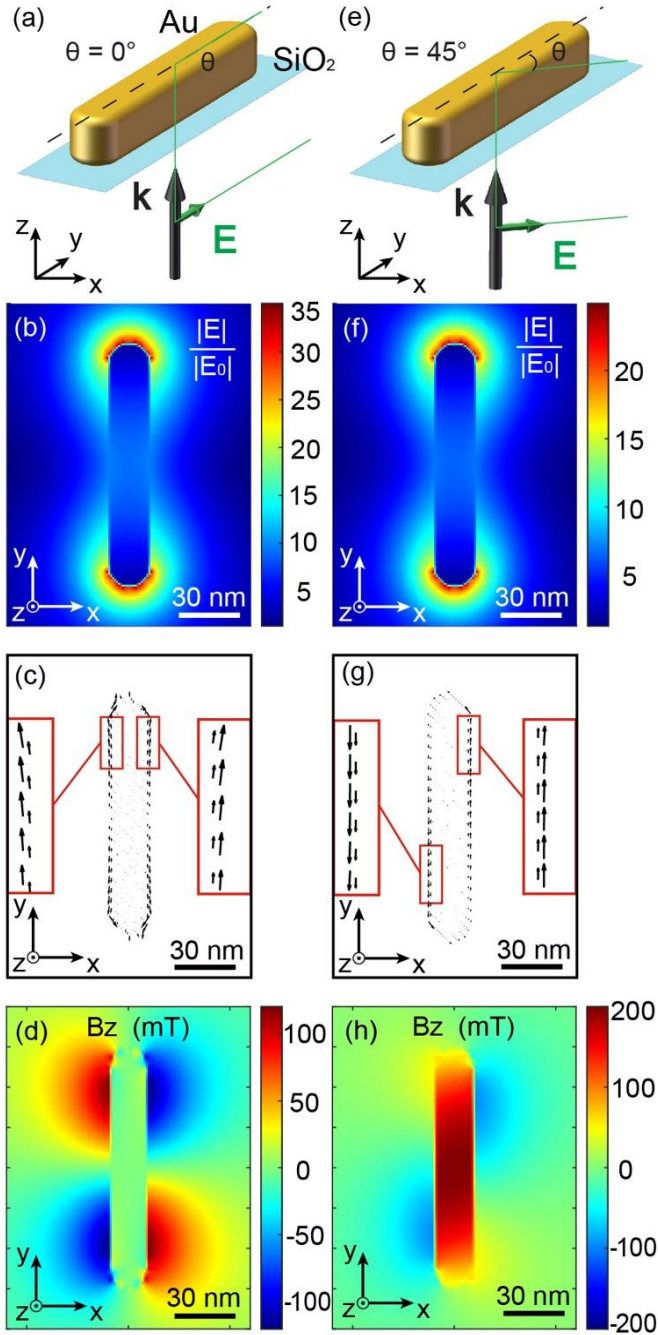

FIG. 3. IFE in a gold nanorod under linear polarization. Generation of drift currents and stationary magnetic fields in a gold nanorod excited by linear polarization (a-d) along the antenna long axes (θ=0°) and (e-h) at 45° from the x-axis. (a, e) Schematic representation of the gold nanorod excited by linear polarization at θ=0° and 45° from the antenna long axes, respectively, at the wavelength of 800 nm and for an intensity of $10^{12}$ W/cm$^2$. For the two excitation conditions shown in (a, e), the normalized electric field distributions (b, f), associated drift currents (c, g), and generated stationary magnetic fields (d, h) are shown, respectively.

In addition, by rotating the linear incident polarization, we show that the orientation of this magnetic field is reversible. In particular, Fig. 4 plots the value of the z-oriented **B**-field at the center of the antenna for different θ angles of the polarization, θ = 0 being the polarization oriented along the long axis of the nanorod. The magnetic field oscillates between a positive maximum for θ = 45° and a negative minimum for θ = 135°. Hence, **B** is zero for θ = 0, 90 and 180°, as shown in Fig. 3(d). Therefore, it is noteworthy that a linear IFE in a plasmonic antenna can, on-demand, reverse the magnetic field generated by simply rotating the incident polarization. This type of behavior would be extremely useful in the all-magnetic ultrafast manipulation of magnetic domains since two opposite magnetic field orientations are achievable for the same power density and heat generation in the structure.

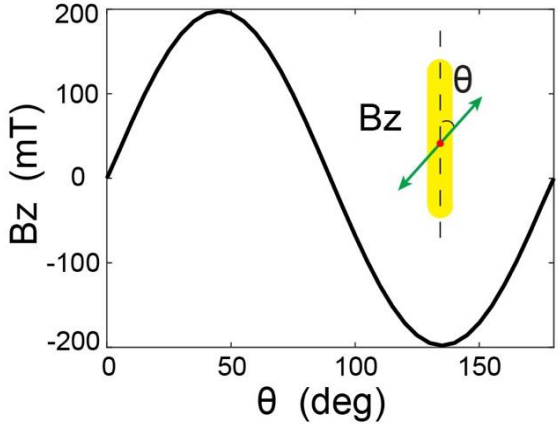

FIG. 4. Magnetic field along z generated by linear IFE at the center of the plasmonic antenna as a function of the incident polarization angle θ.

## IV. SPIN DENSITY AND SUPER CIRCULAR LIGHT

This unexpected physical behavior is surprising because of the similar optical electric field distributions for the different polarization angles used and described in Fig. 3. The explanation is to be found in the dipolar character of this antenna. Indeed, it is known that a gold nanorod behaves like an electric dipole, both in terms of radiation pattern [40] and field lines [41]. Recently, it has also been demonstrated theoretically and experimentally that an electric dipole possesses non-zero local spin densities **s** [Eq. (2)] in the near field [42,43]; that is to say, local distributions of electric fields circularly polarized.

$$\mathbf{s} = \frac{1}{|\mathbf{E_0}|^2}\text{Im}(\mathbf{E}^* \times \mathbf{E}) \qquad (2)$$

With $\mathbf{E_0}$ the electric field of the incoming light.

Fig. 5(a) describes this little-known phenomenon. This figure represents the spin density in a xy transverse plane in the center of the rod for an incident polarization oriented along y. We see that four lobes appear at the four corners of the nanorod with perfectly equal amplitude, with the same sign on the diagonal of the antenna but with opposite signs on the same edge. The different signs of **s** can be understood as light confinements of different helicity. A positive spin density can be associated with a right circular polarization, while a negative density with a left circular polarization. This symmetrical distribution of spin densities shown in Fig. 5(a) is at the origin of the symmetrical drift currents displayed in Fig. 3(c) and the global zero magnetic field in Fig. 3(d).

On the other hand, as observed in Fig. 5(b), changing the angle of the incident polarization θ to 45° modifies the spin density distribution altogether. While it was perfectly symmetrical in the case of θ=0°, the distribution is now completely inhomogeneous, with large lobes along the nanorod (with negative **s**), and two smaller spots at the tips of the antenna (with positive **s**). This dissymmetry induces the asymmetric drift currents presented in Fig. 3(g) and, therefore, the non-zero magnetic field in Fig. 3(h). See Fig. (S2) in supplemental material for the spatial distributions of spin density for different θ angles of the incident polarization.

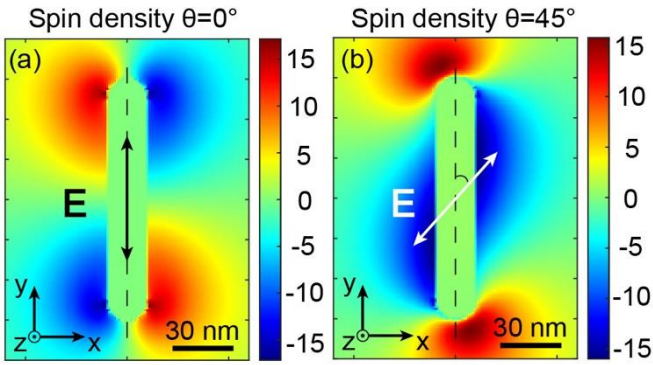

FIG. 5. Near-field spin density of the gold nanorod in a xy transverse plane for linear polarization a) oriented along y and b) oriented at 45° from the long axis of the antenna.

To explain this entirely new physical behavior of linear IFE, it is necessary to decompose the excitation of the plasmonic antenna for an angle θ into its two contributions of θ = 0° and θ = 90°, which we will call the longitudinal and transverse modes, respectively. The total electric field in the near field of the antenna as a function of θ is then written **E**(θ)=**E**$^L$(θ)+**E**$^T$(θ), with **E**$^L$ the electric field of the longitudinal mode and **E**$^T$ that of the transverse mode. The spin density **s** then becomes:

$$\mathbf{s}(\theta) = \text{Im}\left[\left(\mathbf{E}^L(\theta) + \mathbf{E}^T(\theta)\right)^* \times \left(\mathbf{E}^L(\theta) + \mathbf{E}^T(\theta)\right)\right] \quad (3)$$

Which can be decomposed into three terms:

$$\mathbf{s}(\theta) = \text{Im}\left[\left(\mathbf{E}^L(\theta)\right)^* \times \left(\mathbf{E}^L(\theta)\right)\right] + \text{Im}\left[\left(\mathbf{E}^T(\theta)\right)^* \times \left(\mathbf{E}^T(\theta)\right)\right] + 2 \times \text{Im}\left[\left(\mathbf{E}^L(\theta)\right)^* \times \left(\mathbf{E}^T(\theta)\right)\right] \quad (4)$$

Or

$$\mathbf{s}(\theta) = \mathbf{s}^L(\theta) + \mathbf{s}^T(\theta) + \mathbf{s}^{L-T}(\theta) \quad (5)$$

Where **s**$^L$(θ) and **s**$^T$(θ) are the longitudinal and transverse components of **s**(θ), respectively, and **s**$^{L-T}$(θ) is the spin density resulting from mode coupling.

Fig. 6 displays the different spin densities for a 45° angle θ. Fig. 6(a) represents the longitudinal part **s**$^L$(45°) of **s**. As one can see, its distribution is the same as that of Fig. 5(a), but with a lower amplitude due to a lower amount of energy allowed to this mode. Fig. 6(b) depicts the transverse mode **s**$^T$(45°) of **s**, and shows a symmetrical distribution in the four corners of the antenna, just like the longitudinal one, but with a much lower amplitude, due to a lack of resonance. Consequently, **s**$^L$ and **s**$^T$ will not create a global non-zero magnetic field since in these symmetries, the drift currents cancel out. On the other hand, the coupled mode exhibits a distribution of spin density that is also symmetrical but distributed mainly along the nanorod and, remarkably, with higher amplitude than each of the modes taken separately. Hence, the sum of the three spin density modes **s**$^L$, **s**$^T$, and **s**$^{L-T}$ gives the distribution of **s** shown in Fig. 6(d) and identical to Fig. 5(b), at the origin of the asymmetric drift currents [Fig. 3(g)] generating the magnetic field presented in Fig. 3(h). Thus, the coupling between the two longitudinal and transverse modes of this plasmonic dipole antenna allows generating an IFE via a linear polarized light. As well, since the energy splits into the longitudinal and transverse modes by a factor cos(θ) and sin(θ), respectively, the coupling between these two modes is proportional to sin(2θ) (Eq. (4)) and, thus, is maximum for an angle θ of 45° and 135°, as shown in Fig. 4.

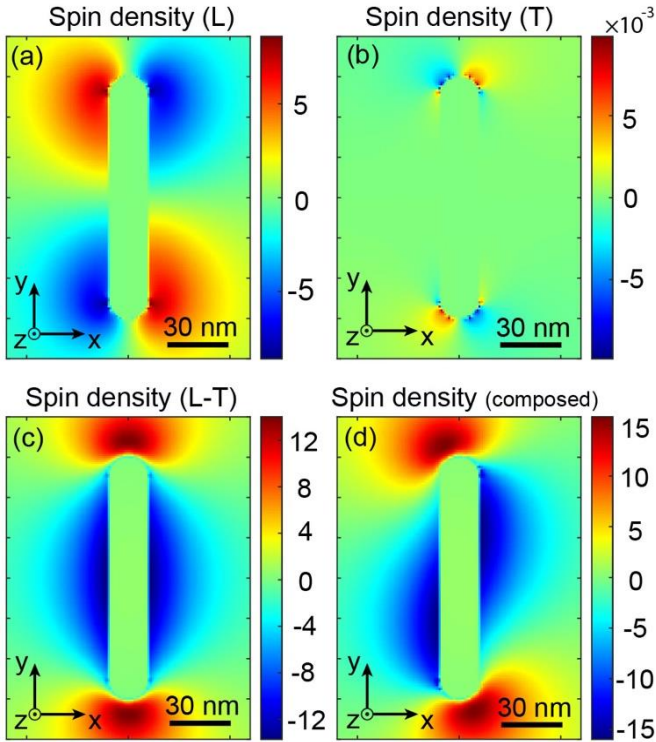

FIG. 6. Different contributions of the nanorod modes in the local spin density. Spin density due to the contribution of the a) longitudinal, b) transverse, c) coupled modes. d) Sum of the spin densities of the different modes represented in a-c.

As a final note, and although it is beyond the scope of this article, it is interesting to note that the spin densities generated by the nanorod are always higher than those created in the far-field. Indeed, for a circularly polarized plane wave, the spin density calculated by Eq. (2) can only take a value of -1 or 1, depending on the helicity of the light. Here, owing to the resonating nature of this antenna, the spin density can be increased up to values of -15 to 15, depending on the polarization angle incident on the nanostructure [Figs. 4,5 and (S2)], thus introducing the concept of super circular light, similar to the previously developed ideas of super chiral light [44].

**V. CONCLUSION**

In conclusion, for the first time since its discovery, we have demonstrated that a linear polarization of light could generate an inverse Faraday effect. While this magneto-optical phenomenon was believed to be generated only by circular or elliptical polarization, we have shown that this postulate was no longer valid by manipulating light at the nanoscale with a plasmonic nanostructure. The demonstration of this new physical effect originates in the local manipulation of the optical nano-antenna's spin density (local polarization state of the light). In particular, the use of a gold nanorod excited by a linear polarization oriented at 45° of its principal axes generates locally a spin density associated with a super circular light, creating, in turn, drift currents via IFE and thus the magnetization of the antenna. The magnetic field produced is then reversible on demand simply by changing the polarization angle with respect to the nanorod. Its intensity is up to 25 times higher than a gold nanoparticle excited by a circular polarization of the same power and frequency. The results presented in this work go beyond the simple demonstration of a new physical effect; they also open exciting perspectives for manipulating and controlling magnetic processes at ultrashort time scales. Indeed, using an IFE in plasmonic nanostructures is nowadays the only technique allowing the generation of intense and ultra-short magnetic fields at the nanoscale. Thus, a linear IFE will allow the ultrafast manipulation of magnetic domains in the near field for the same excitation polarization, energy density and heat generation, simply by changing the angle of the incident polarization on the plasmonic nano-antenna.

**ACKNOWLEDGMENTS**

We acknowledge the financial support from the Agence national de la Recherche (ANR-20-CE09-0031-01), from the Institut de Physique du CNRS (Tremplin@INP 2020) and the China Scholarship Council.